\documentclass[aps,pre,twocolumn,showpacs]{revtex4}
\usepackage{epsfig}
\usepackage{times}
\begin{document}

\title{Cooperation in spatial Prisoner's Dilemma with two types of players \\ for increasing number of neighbors}

\author{Gy\"orgy Szab\'o and Attila Szolnoki}
\affiliation {Research Institute for Technical Physics and
Materials Science, P.O. Box 49, H-1525 Budapest, Hungary}

\begin{abstract}
We study a spatial two-strategy (cooperation and defection) Prisoner's Dilemma game with two types ($A$ and $B$) of players located on the sites of a square lattice. The evolution of strategy distribution is governed by iterated strategy adoption from a randomly selected neighbor with a probability depending on the payoff difference and also on the type of the neighbor. The strategy adoption probability is reduced by a pre-factor ($w < 1$) from the players of type $B$. We consider the competition between two opposite effects when increasing the number of neighbors ($k=4$, 8, and 24). Within a range of the portion of influential players (type $A$) the inhomogeneous activity in strategy transfer yields a relevant increase (dependent on $w$) in the density of cooperators. The noise-dependence of this phenomenon is also discussed by evaluating phase diagrams.
\end{abstract}


\maketitle

The investigation of the spatial evolutionary Prisoner's Dilemma
(PD) games expands progressively since Nowak and May \cite{nowak_ijbc93} reported the maintenance of cooperative behavior among selfish players. In these models the PD game \cite{gintis_00} represents a pair interaction between two players who can either cooperate ($C$) or defect ($D$) and their income depends on both choices in a way forcing both rational (selfish) players to choose $D$ while they would share equally the maximum total payoff for mutual cooperation. 

In the first model the players are located on a square lattice, they can follow one of the two pure strategies ($D$ and $C$) and their income comes from PD games with the neighbors. During a synchronized strategy update the players adopt the strategy from the neighbor receiving the highest score. After this pioneering work many modified versions of the original model have been suggested and studied (for recent surveys see \cite{nowak_06,szabo_pr07}). Let us mention only a couple of examples: in some models larger set of strategies was used \cite{lindgren_pd94,frean_prslb94,hauert_jtb98}, in the evolutionary rules noises \cite{nowak_ijbc94,blume_geb03,szabo_pre98} were introduced that can help the cooperative behavior \cite{vukov_pre06,perc_njp06a,perc_njp06c}, and the spatial structure was also extended by locating the players on different graphs giving a better description about the connections in human societies \cite{abramson_pre01,kim_pre02,masuda_pla03,duran_pd05}. In the last years the concept of interaction and learning graphs are distinguished \cite{ohtsuki_prl07,ohtsuki_jtb07b,wu_pre07} and the research of the co-evolution of strategy distribution and these graph has also become a promising topic \cite{pacheco_prl06,pacheco_jtb06,pacheco_jtb08}. It is found, furthermore, that different types of personality \cite{dall_el04,perc_pre08} and inhomogeneous activity in the strategy adoption can also support cooperation \cite{kim_pre02,wu_cpl06,wu_pre06,guan_epl06} particularly if some distinguished players have higher influence to spread their strategies \cite{szolnoki_epl07,szolnoki_epjb08}. 

In the latter case the relevant increase in the frequency of cooperators is related to a phenomenon described previously by Santos {\it et al}. \cite{santos_prl05,santos_prslb06} who studied evolutionary PD games on scale-free networks with an evolutionary rule exploiting the high income for players (called hubs) who have a large number of neighbors. As a result, on the scale-free networks the strategy of hubs become an example to be followed by their neighborhood. Thus, the hubs as influential players face the consequence of the imitation of their own strategy that increases (decreases) the income of cooperative (defective) hubs. After a short transient process this phenomenon favors the spreading of cooperation because the influential players can also adopt strategy from each other for suitable connectivity structures. Evidently, in the absence of links between influential players the mentioned mechanism cannot help cooperators to beat defectors \cite{santos_prl05,santos_prslb06,rong_pre07}. Recent studies \cite{szolnoki_pa08,szolnoki_epjb08} have indicated that the presence of linked influential players on scale-free graphs can efficiently promote cooperation (even for normalized payoffs) if the capability of strategy spreading differs from player to player. These results raise many interesting questions about the impact of the size of neighborhood on the frequency of cooperators for inhomogeneous activity in the strategy transfer.

In the present work we study the competition between two opposite effects emerging if the average number of neighbors is increased. On one hand, the above described mechanism (supporting the spreading of cooperation for inhomogeneous strategy transfer capability) is enhanced when choosing larger and larger $k$. On the other hand, the increase of the number of neighbors, $k$, is beneficial for defectors on regular networks \cite{kirchkamp_jebo00,ifti_jtb04,santos_prslb06,tang_epjb06,ohtsuki_n06,
szamado_jtb08,hatzopoulos_pre08}. Here it is worth mentioning that the mean-field approximation (predicting the extinction of cooperators in the evolutionary PD games \cite{nowak_06,szabo_pr07}) gives a simple explanation of this phenomenon. The scope of the present paper is to explore the impact of these two opposite effects by comparing results obtained for three different sizes of neighborhood. More precisely, the studied types of neighborhood are the von Neumann neighborhood including only the nearest neighbors ($k=4$), the Moore neighborhood with nearest and next-nearest neighbors ($k=8$), and the case of $k=24$ where players within a $5 \times 5$ box of sites are neighbors of the central player (self-interaction is excluded). Monte Carlo (MC) simulations are used to study systematically the effects of payoff, number of neighbors, and inhomogeneous capability of strategy transfer (for a fixed noise level) on the average number of cooperators.

For these evolutionary PD game models two types of players ($n_x=A$ or $B$) are located on the sites $x$ of a square lattice with a concentration of $\nu$ and $(1-\nu)$ and their random initial distribution remains unchanged (quenched) during the simulations. The income of player $x$ comes from one-shot PD games with her neighbors, that is, 
\begin{equation}
P_x=\sum_{y \in \Omega_x} {\bf s}_x \cdot {\bf A} {\bf s}_y
\label{eq:sumpayoff}
\end{equation}
where the sum runs over all neighboring sites ($\Omega_x$) of player $x$, the payoff matrix is defined as suggested by Nowak and May \cite{nowak_ijbc93},
\begin{equation}
\label{eq:pom}
{\bf A}=\left( \matrix{0 & b \cr
                       0 & 1 \cr} \right)\;, \;\; 1 < b < 2 \;, 
\end{equation} 
and the defective and cooperative strategies are denoted by unit vectors as
\begin{equation}
\label{eq:sx}
{\bf s}_x=D=\left( \matrix{1 \cr
                           0 \cr} \right)\;\;\; 
                       \mbox{or} \;\;\; 
C=\left( \matrix{0 \cr
                 1 \cr} \right) \;.
\end{equation} 
The evolution of strategy distribution is governed by random sequential strategy update representing strategy adoption from a randomly chosen site $y$ to one of its neighbors $x$ with a probability
\begin{equation}
W(s_x \leftarrow s_y)=w_y\,\frac{1}{1+\exp[(P_x-P_y)/k K]}\,\,
\label{eq:prob}
\end{equation}
dependent on the difference of normalized payoffs (e.g. $(P_x-P_y)/k$) for later convenience of comparisons. For this strategy adoption probability  the meaning of the parameter $K$ is analogous to the temperature as introduced in the kinetic Ising model and characterizes the magnitude of payoff noise affecting the decision of player $x$ \cite{blume_geb03,szabo_pre98}. The multiplicative factor $w_y$ denotes the strategy transfer capability of player $y$,
\begin{equation}
\label{eq:wobt}
w_{y}= \cases {1, &if $n_y=A$ \cr
                w, &if $n_y=B$ \cr }\;\;\;,\;\; 0<w < 1 \;.
\end{equation} 
In this notation players of type $A$ represent those individuals who can easily convince their neighbors to adopt the strategy they are just following. This personal feature can be related to age, reputation, {\it etc}. 

For all the three cases studied here the simulations are performed on an $L \times L$ square of sites with periodic boundary conditions. The evolution of the spatial distribution of the $C$ and $D$ strategies starts from an uncorrelated initial state where cooperators and defectors are present with the same probability. When repeating the above described elementary steps the system develops into a final stationary state characterized by the average density of cooperators ($\rho$). After a suitable relaxation time $t_r$ $\rho$ is determined by averaging the density of cooperators over a time $t_a$. Typical (maximum) value of parameters used in our simulations are the following: $L=400$ ($1600$); $t_r \simeq t_a=10^4$  ($10^6$) MCS [during one MC step (MCS) each player has a chance once on average to adopt one of the neighboring strategies]. Pronounced long relaxations were observed at the large noise limit.

Before discussing the behaviors of the above systems we briefly recall some general features of the homogeneous system ($\nu =0$ or 1) \cite{szabo_pre05}. The average (total) payoff increases monotonously with $\rho$ independent of the initial strategy distribution. Furthermore, in each homogeneous system the value of $\rho$ decreases monotonously from 1 to 0 if $b$ is increased within a region of $b$ ($b_{c1}^{(k)} (K)< b < b_{c2}^{(k)}(K)$) where the strategies $C$ and $D$ coexist. For all the three types of neighborhoods $b_{c1}^{(k)}(K)$ ($b_{c2}^{(k)}(K)$) tends to 1 from below (above) if $K \to \infty$. In other words, in the strong noise limit ($K \to \infty$) the systems reproduce the behavior of the mean-field model, that is $\rho$ drops suddenly from $1$ to $0$ at $b=1$. In the opposite case ($K \to 0$) the limit values of $b_{c1}^{(k)}$ and $b_{c2}^{(k)}$ depend on $k$. When decreasing $K$ the upper critical value of $b$ tends monotonously to a value ($b_{c1}^{(k)}(0)$ larger than 1 if $k=8$ or $24$. On the contrary, for $k=4$, the function $b_{c2}^{(4)}(K)$ has a local maximum at $K=K_{opt} \simeq 0.08$  ($b_{c2}^{(4)}(K=K_{opt}) \simeq 1.08$) and approaches 1 if $K \to 0$.

\begin{figure}[ht]
\centerline{\epsfig{file=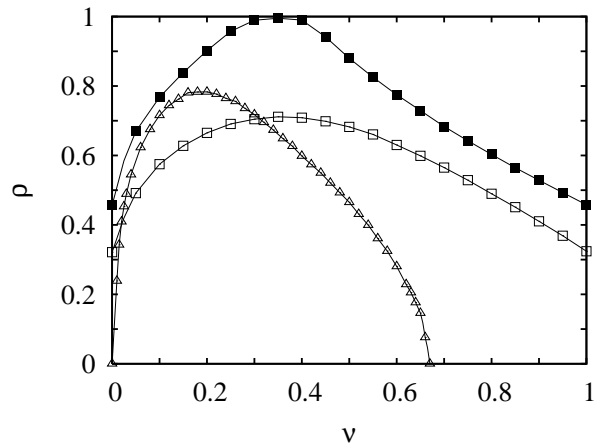,width=8cm}}
\caption{\label{fig1}Density of cooperators as a function of the portion of $A$ players if $b=1.05$, $K=0.1$ and $w=0.1$ for three different neighborhoods: $k=4$ (open squares); 8 (closed squares); and 24 (closed triangles).}
\end{figure}

In the light of the above results we first study the density of cooperators ($\rho$) when varying the portion of players of type $A$ for fixed values of payoff ($b$), strategy transfer capability ($w$), and noise ($K$). The latter was chosen to present optimal cooperation for $k=4$ system (i.e., $K \cong K_{opt}$). The MC data are compared in Fig.~\ref{fig1} for the three types of neighborhood. 
For the sake of comparison, we selected such a high value of $b$, which prevents cooperation in the homogeneous model of $k=24$. Figure \ref{fig1} shows that the highest density of cooperators can be observed (at $\nu=0$ or 1) for $k=8$ where the overlapping triangles in the connectivity structure support the spreading (maintenance) of cooperation as discussed in \cite{vukov_pre06}. The further increase of $k$, however, yields a decrease in both $\rho$ and $b_{c2}^{(k)}$ \cite{santos_prslb06} tending to the behavior of mean-field model. This is the reason why cooperators become extinct in the final stationary state for the homogeneous system at $k=24$.

Figure \ref{fig1} demonstrates clearly the existence of an optimum composition (defined by the maximum in $\rho$) of the players $A$ and $B$. The presence of distinguished players results in a relatively higher impact on cooperation level for larger $k$. In agreement with the expectations, the more neighbors the players have, the smaller portion of influential players (type $A$) are capable to achieve the highest increase in $\rho$. The resultant asymmetry can be observed in the function $\rho(\nu)$ for $k=24$. For the largest neighborhood our simulations have clearly indicated that cooperators can remain alive only within a region of $\nu$ with boundaries dependent on $w$ and $K$. It is expected, that this region, $\nu_1(w,K) < \nu < \nu_2(w,K)$, shrinks if we increase $k$ further.

\begin{figure}[ht]
\centerline{\epsfig{file=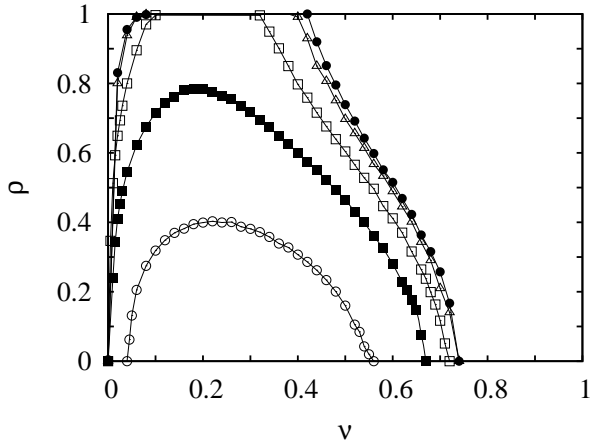,width=8cm}}
\caption{\label{fig2}Density of cooperators {\it vs}. $\nu$ for five different values of the reduced strategy transfer capability ($w=0.01$, 0.02, 0.05, 0.1, and 0.2 from top to bottom) at $b=1.05$, $k=24$, and $K=0.1$.}
\end{figure}

As the largest effect is found for the largest neighborhood, henceforth our attention will be focused on the system of $k=24$. Figure \ref{fig2} illustrates the increase of the density of cooperators when varying the composition of players $A$ and $B$ for several values of $w$ at a fixed payoff and noise level. 
When the difference is small, typically when $1/w \le 2$, the cooperators cannot remain alive at the given payoffs and noise independently of the actual composition of $A$ and $B$ players. If the ratio $1/w$ is increased then the cooperators can survive within the above mentioned region of $\nu$. This interval becomes wider and wider while the maximum value of $\rho$ increases monotonously until reaching its saturation value ($\rho=1$). Consequently, we can observe four subsequent transitions in Fig.~\ref{fig2} if $\nu$ increased for sufficiently high values of the ratio $1/w$. Apparently the density of cooperators tends to a limit profile if $1/w \to \infty$. We have to emphasize that the rigorous analysis of the asymptotic behavior becomes difficult because the transient time increases with the ratio $1/w$ particularly at small values of $\nu$.

\begin{figure}[ht]
\centerline{\epsfig{file=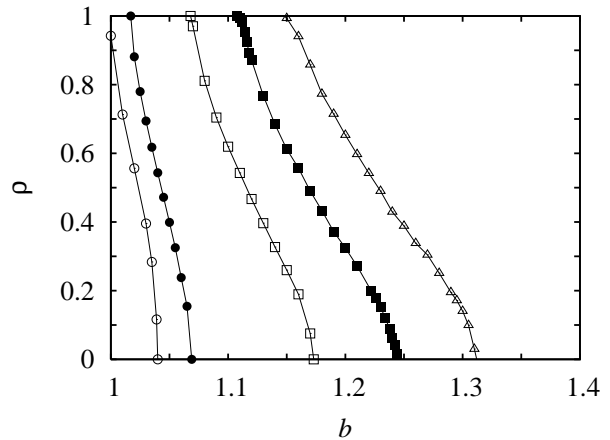,width=8cm}}
\caption{\label{fig3}Density of cooperators as a function of $b$ for different values of $1/w$ ($w=1$, 0.2, 0.05, 0.02m and 0.005 from left to right) at fixed noise level ($K=0.1$), composition ($\nu=0.2$), and neighborhood ($k=24$).}
\end{figure}

We have also studied the effect of the variation of $w$ on the cooperation level at different payoffs ($b$). To avoid additional effects the
noise level is fixed at a composition ($\nu=0.2$) close to its optimum value. The results, summarized in Fig.~\ref{fig3},
illustrates that the curves $\rho(b)$ shift to larger $b$ values if the ratio $1/w$ is increased. (For comparison, the left curve shows the results obtained in the homogeneous system). 
The plotted results refer to a shift proportional to $\ln(1/w)$. Due to the above mentioned increasing run time if we choose larger values of $1/w$, the more rigorous (numerical) confirmation of this trend goes beyond the scope of the present work. Instead of it we have concentrated on the effect of noise for the two extreme neighborhoods ($k=4$ and $24$) at a fixed portion of players $A$ and $B$. For this purpose we have performed systematic MC simulations to determine the critical values $b_{c1}$ and $b_{c2}$ for a fixed ratio of strategy transfer capability ($1/w=50$).

\begin{figure}[ht]
\centerline{\epsfig{file=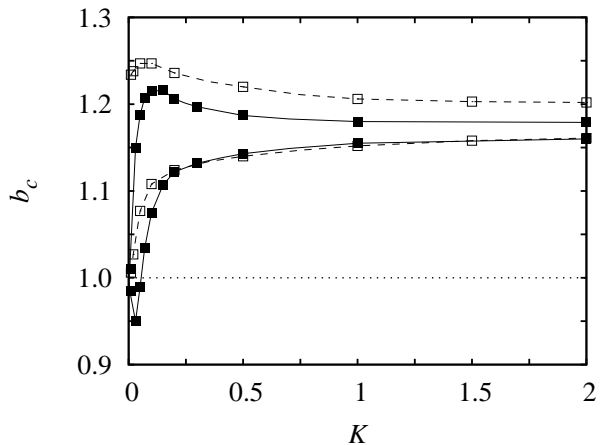,width=8cm}}
\caption{\label{fig4}The upper and lower critical values of $b$ for $k=24$ and $\nu=0.2$ (open squares connected with dashed lines). Results for $k=4$ and $\nu=0.5$ are denoted by closed squares (connected with solid lines) at $w=0.02$. The dotted line illustrates the prediction of mean-field approximation in the homogeneous system ($b_{c1}^{\rm (mf)}=b_{c2}^{\rm (mf)}=1$ for arbitrary $K$).}
\end{figure}

Figure \ref{fig4} can be interpreted as a phase diagram where the connected data represent phase boundaries. Between the upper and lower critical points strategies $C$ and $D$ coexist. Above (below) this region only defectors (cooperators) remain alive in the final stationary states. For both cases the system behavior is not affected by the spatial inhomogeneities in the low noise limit, in agreement with the previous results \cite{szolnoki_epl07}. In other words, the relevant improvement in the maintenance of cooperation appears in the noisy systems even for the limit $K \to \infty$. In contrary to the prediction of mean-field theory the present data indicate clearly that cooperators and defectors can coexist within a region of $b$ if $K$ goes to infinity, that is, $b_{c1}$ and $b_{c2}$ tends to two distinct limit values for both types of neighborhood. This latter feature has already been confirmed qualitatively by the pair approximation for $k=4$ \cite{szolnoki_epl07}. Notice, furthermore, that the larger neighborhood yields a larger increase in the value of $b_{c1}$ and $b_{c2}$ when applying optimum composition of players $A$ and $B$ for both systems.  

In summary, within the framework of evolutionary PD games, the present investigation of the effect of the inhomogeneous strategy transfer capability on the cooperative behavior has indicated a relevant increase in the density of cooperators if the fraction of influential players was close to the optimum value dependent on the number of neighbors (range of interaction) if two types of strategy transfer capability (represented by the players $A$ and $B$) are distinguished. It is found that the larger neighborhood with a less fraction of influential players (type $A$) can be more beneficial for the whole system due to the imitation mechanism rewarding (punishing) cooperation (defection) for the influential players. The improvement of cooperation increases with the ratio of strategy transfer capability ($1/w$) between players of type $A$ and $B$. Furthermore, the maintenance of cooperation is supported remarkably by the mentioned effect in the high noise limit where the region of coexistence is broadened and shifted to higher values of $b$. For small fraction of players $A$ one can think that the competition between the influential players surrounded by their followers can be characterized by an effective (rescaled) payoff matrix favoring cooperation (as it appears on evolving networks \cite{pacheco_prl06}) while their competitions are disturbed by those players of type $B$ who do not belong to the neighborhood of any influential player. These latter $B$ players can mediate an interaction between the influential players and/or preserve the defective behavior. Further research is requested to clarify the relevance of these opposite effects.

\begin{acknowledgments}

This work was supported by the Hungarian National Research Fund
(Grant No. K-73449).

\end{acknowledgments}


\end{document}